\documentclass[namedreferences]{SolarPhysics}
\usepackage[optionalrh,solaenum]{spr-sola-addons} 
\usepackage{subfig}
\usepackage{caption}
\usepackage{multirow}
\usepackage{graphicx}        
 \usepackage{color}           
\usepackage{url}             
 \usepackage{amssymb}
 
\begin{document}
\begin{article}
\begin{opening}
\title{Magnetic Connectivity between Active Regions 10987, 10988, and 10989 by Means of Nonlinear Force-Free Field Extrapolation}
\author{Tilaye~\surname{Tadesse}$^{1}$\,$^{2}$,
         T.~\surname{Wiegelmann}$^{1}$, and
        B.~\surname{Inhester}$^{1}$, and
        A.~\surname{Pevtsov}$^{3}$
       }
\runningauthor{T.~Tadesse et al.}
\runningtitle{Nonlinear force-free field extrapolation in spherical geometry}

   \institute{$^{1}$ Max Planck Institut f\"{u}r Sonnensystemforschung, Max-Planck Str. 2, D--37191 Katlenburg-Lindau, Germany
                     email: \url{tilaye.tadesse@gmail.com}, email: \url{wiegelmann@mps.mpg.de}, email: \url{inhester@mps.mpg.de}\\
                $^{2}$Addis Ababa University, College of Education, Department of Physics
 Education, Po.Box 1176, Addis Ababa, Ethiopia \\
              $^{3}$ National Solar Observatory, Sunspot, NM 88349, U.S.A.
                     email: \url{apevtsov@nso.edu} \\
             }

\begin{abstract}
Extrapolation codes for modelling the magnetic field in the corona in cartesian geometry
do not take the curvature of the Sun's surface into account and can only be applied to
relatively small areas, \textit{e.g.}, a single active region. We apply a method for nonlinear
force-free coronal magnetic field modelling of photospheric vector magnetograms in spherical
geometry which allows us to study the connectivity between multi-active regions.
We use vector magnetograph data from the Synoptic Optical Long-term Investigations of the
Sun survey (SOLIS)/Vector Spectromagnetograph(VSM) to model the coronal magnetic field, where
we study three neighbouring magnetically connected active regions (ARs: 10987, 10988, 10989)
observed on 28, 29, and 30 March 2008, respectively. We compare the magnetic field topologies
and the magnetic energy densities and study the connectivities between the active regions(ARs).
We have studied the time evolution of magnetic field over the period of three days and found no
major changes in topologies as there was no major eruption event. From this study we have concluded
that active regions are much more connected magnetically than the electric current.

\end{abstract}
\keywords{Magnetic fields · Photosphere · Corona}
\end{opening}
\section{Introduction}
     \label{S-Introduction}
In order to model and understand the physical mechanisms underlying the various activity phenomena that can be
observed in the solar atmosphere, like, for instance, the onset of flares and coronal mass ejections, the stability
of active region, and to monitor the magnetic helicity and free magnetic energy, the magnetic
field vector throughout the atmosphere must be known. However, routine measurements of the solar magnetic field
are mainly carried out in the photosphere. The magnetic field in the photosphere is measured using the
Zeeman effect of magnetically-sensitive solar spectral lines. The problem of measuring the coronal field and
its embedded electrical currents thus leads us to use numerical modelling to infer the field strength in the
higher layers of the solar atmosphere from the measured photospheric field. Except in eruptions, the magnetic
field in the solar corona evolves slowly as it responds to changes in the surface field, implying that the
electromagnetic Lorentz forces in this low-$\beta$ environment are relatively weak and that any electrical
currents that exist must be essentially parallel or antiparallel to the magnetic field wherever the field is not negligible.

Due to the low value of the plasma $\beta$ (the ratio of gas pressure to magnetic pressure), the solar corona
is magnetically dominated \cite{Gary}. To describe the equilibrium structure of the static coronal magnetic field when non-magnetic
forces are negligible, the force-free assumption is appropriate:
\begin{equation}
   (\nabla \times\textbf{B})\times\textbf{B}=0 \label{one}
\end{equation}
\begin{equation}
    \nabla \cdot\textbf{B}=0 \label{two}
 \end{equation}
\begin{equation}
    \textbf{B}=\textbf{B}_{\textrm{obs}} \quad \mbox{on photosphere} \label{three}
 \end{equation}
 where $\textbf{B}$ is the magnetic field and $\textbf{B}_{\textrm{obs}}$ is measured vector field on the
 photosphere. Equation~(\ref{one}) states that the Lorentz force vanishes (as a consequence of
 $\textbf{J}\parallel \textbf{B}$, where $\textbf{J}$ is the electric current density) and Equation~(\ref{two})
 describes the absence of magnetic monopoles.

 The extrapolation methods based on this assumption 
are termed nonlinear force-free field extrapolation
 \cite{Sakurai81,Amari97,Amari99,Amari,wu90,cuperman91,demoulin92,Inhester06,mikic94,Roumeliotis,yan00,valori05,Wiegelmann04,Wheatland04,Wheatland:2009,Wheatland:2010,Amari:2010}.
 For a more complete review of existing methods for computing nonlinear force-free coronal magnetic fields, we refer
 to the review papers by \inlinecite{Amari97}, \inlinecite{Schrijver06}, \inlinecite{Metcalf}, and
 \inlinecite{Wiegelmann08}. \inlinecite{Wiegelmann:2006} has developed a code for the self-consistent computation of the
 coronal magnetic fields and the coronal plasma that uses non-force-free MHD equilibria.

The magnetic field is not force-free in the photosphere, but becomes force-free roughly 400 km above the photosphere
\cite{Metcalf:1995}. Furthermore, measurement errors, in particular for the transverse field components (i.e., perpendicular
to the line of sight of the observer), would destroy the compatibility of a magnetogram with the condition of being
force-free. One way to ease these problems is to preprocess the magnetogram data as suggested by \inlinecite{Wiegelmann06sak}.
The preprocessing modifies the boundary values of  $\textbf{B}$ within the error margins of the measurement in such
a way that the moduli of force-free integral constraints of Molodenskii \cite{Molodenskii74} are minimized. The resulting
boundary values are expected to be more suitable for an extrapolation into a force-free field than the original values.

In the present work, we use a larger computational domain which accommodates most of the connectivity within the coronal
region. We also take the uncertainties of measurements in vector magnetograms into account as suggested in
\inlinecite{DeRosa}.
We apply a preprocessing procedure to SOLIS data in spherical geometry \cite{Tilaye:2009} by taking account of the curvature
of the Sun's surface. For our observations, performed on 28, 29, and 30 March 2008, respectively, the large field of
view contains three active regions (ARs: 10987, 10988, 10989).

The full inversion of SOLIS/VSM magnetograms yields the magnetic filling factor for each pixel, and it also corrects for
magneto-optical effects in the spectral line formation. The full inversion is performed in the framework of Milne-Eddington
model (ME)\cite{Unno:1956} only for pixels whose polarization is above a selected threshold. Pixels with polarization below
threshold are left undetermined.
These data gaps represent a major difficulty for existing magnetic field extrapolation schemes. Due to the large
area of missing data in the example treated here in this work, the reconstructed field model obtained must be treated with some
caution. It is very likely that the field strength in the area of missing data was small because the inversion procedure, which
calculates the surface field from the Stokes line spectra, abandons the calculation if the signal is below a certain threshold.
The magnetic field in the corona, however, is dominated by the strongest flux elements on the surface, even if they occupy only
a small portion of the surface. We are therefore confident that these dominant flux elements are accounted for in the surface
magnetogram, so that the resulting field model is fairly realistic. At any rate, it is the field close to the real, which can
be constructed from the available sparse data. Therefore, we use a procedure which allows us to incorporate measurement error and treat
regions with lacking observational data as in \inlinecite{Tilaye:2010}. The technique has been tested in cartesian geometry in
\inlinecite{Wiegelmann10} for synthetic boundary data.
\section{Optimization Principle in Spherical Geometry}
\inlinecite{Wheatland00} have proposed the variational principle to be solved iteratively which minimizes Lorentz
forces (\ref{one}) and the divergence of magnetic field (\ref{two}) throughout the volume of interest, $V$.
Later on the procedure has been improved by \inlinecite{Wiegelmann04} for cartesian geometry in a such way that it can only
uses the bottom boundary on the photosphere as an input. Here we use optimization approach for the functional
$(\mathcal{L}_\mathrm{\omega})$ in spherical geometry \cite{Wiegelmann07,Tilaye:2009} and iterate $ \textbf{B}$
to minimize $\mathcal{L}_\mathrm{\omega}$. The modification concerns the input bottom boundary field $\textbf{B}_{\textrm{obs}}$
which the model field $\textbf{B}$ is not forced to match exactly but we allow deviations of the order of
the observational errors. The modified variational problem is \cite{Wiegelmann10,Tilaye:2010}:
 \begin{displaymath} \textbf{B}=\textrm{argmin}(\mathcal{L}_{\omega})
\end{displaymath}
 \begin{equation}\mathcal{L}_\mathrm{\omega}=\mathcal{L}_{\textrm{f}}+\mathcal{L}_{\textrm{d}}+\nu\mathcal{L}_{\textrm{photo}} \label{four}
\end{equation}
\begin{displaymath} \mathcal{L}_{\textrm{f}}=\int_{V}\omega_{\textrm{f}}(r,\theta,\phi)B^{-2}\big|(\nabla\times\textbf{B})\times
\textbf{B}\big|^2  r^2\sin\theta dr d\theta d\phi
\end{displaymath}
\begin{displaymath}\mathcal{L}_{\textrm{d}}=\int_{V}\omega_{\textrm{d}}(r,\theta,\phi)\big|\nabla\cdot\textbf{B}\big|^2
  r^2\sin\theta dr d\theta d\phi
\end{displaymath}
\begin{displaymath}\mathcal{L}_{\textrm{photo}}=\int_{S}\big(\textbf{B}-\textbf{B}_{\textrm{obs}}\big)\cdot\textbf{W}(\theta,\phi)\cdot\big(
\textbf{B}-\textbf{B}_{\textrm{obs}}\big) r^{2}\sin\theta d\theta d\phi
\end{displaymath}
where $\mathcal{L}_\mathrm{\textrm{f}}$ and $\mathcal{L}_\mathrm{\textrm{d}}$ measure how well the force-free Equations (\ref{one}) and
divergence-free (\ref{two}) conditions are fulfilled, respectively. $\omega_{\textrm{f}}(r,\theta,\phi)$ and $\omega_{\textrm{d}}(r,\theta,\phi)$
are weighting functions for force-free term and divergence-free term, respectively and identical for this study. The third integral, $\mathcal{L}_{\textrm{photo}}$, is a surface integral over the photosphere which
serves to relax the field on the photosphere towards a force-free solution without too much deviation from the original
surface field data, $\textbf{B}_{\textrm{obs}}$. In this integral, $\textbf{W}(\theta,\phi)=\textrm{diag}(w_{\textrm{radial}},w_{\textrm{trans}},w_{\textrm{trans}})$
is diagonal matrix which gives different weights for observed surface field components depending on its relative
accuracy in measurement. In this sense, lacking data is considered most inaccurate and is taken account of by
setting $W(\theta,\phi)$ to zero in all elements of the matrix.
\begin{figure}
\centering
\includegraphics[bb=30 20 575 382,clip,width=1.0\textwidth]{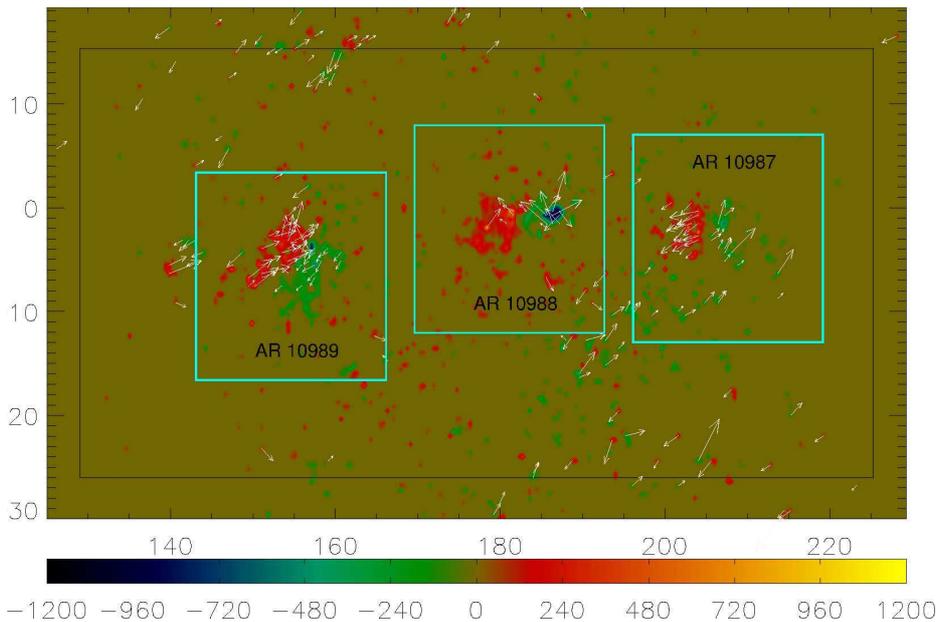}
\caption{Surface contour plot of radial magnetic field component and vector field plot of
transverse field with white arrows. The color coding shows $B_{r}$ on the photosphere. The vertical and horizontal
axes show latitude, $\theta$(in degree) and longitude, $\phi$(in degree) on the photosphere. In the area
coloured in olive, field values are lacking. The region inside the black box corresponds to the physical domain where
the weighting function is unity and the outside region is the buffer zone where it declines to zero. The blue boxes indicate
the domains of the three active regions.}
\label{fig1}
\end{figure}

We use a spherical grid $r$, $\theta$, $\phi$ with $n_{r}$, $n_{\theta}$, $n_{\phi}$ grid points in the
direction of radius, latitude, and longitude, respectively. In the code, we normalize the magnetic field with the average
radial magnetic field on the photosphere and the length scale with a solar radius for numerical reason. Figure~\ref{fig1}
shows a map of the radial component of the field as color-coded and the transverse magnetic field depicted as white arrows.
For this particular dataset, about $86\%$ of the data pixels are undetermined.
The method works as follows:
\begin{itemize}
      \item We compute an initial source surface potential field in the computational domain from
$\textbf{B}_{\textrm{obs}}\cdot\hat{r}$, the normal component of the surface field at the photosphere at $r = 1R_\mathrm{\odot}$.
The computation is performed by assuming that a currentless ($\textbf{J}=0$ or
$\nabla\times\textbf{B}=0$) approximation holds between the photosphere and some spherical surface $S_{s}$
(source surface where the magnetic field vector is assumed radial). We compute the solution of this boundary-value
problem in a standard form of spherical harmonics expansion.
      \item We minimize $\mathcal{L}_{\omega}$(Equations \ref{four}) iteratively. The model magnetic field
      $\textbf{B}$ at the surface is gradually driven towards the observed field $\textbf{B}_{\textrm{obs}}$ while the field
      in the volume $V$ relaxes to force-free. If the observed field, $\textbf{B}_{\textrm{obs}}$, is inconsistent,
      the difference $\textbf{B}-\textbf{B}_{\textrm{obs}}$ remains finite depending in the control parameter $\nu$.
      At data gaps in $\textbf{B}_{\textrm{obs}}$, the respective field value is automatically ignored.
      \item The iteration stops when $\mathcal{L}_{\omega}$ becomes stationary as
   $\Delta \mathcal{L}_{\omega}/\mathcal{L}_{\omega}<10^{-4}$, $\Delta \mathcal{L}_{\omega}$ is the decrease of
   $\mathcal{L}_{\omega}$ during an iterative steps.
   \item A convergence to $\mathcal{L}_{\omega}=0$ yields a perfect force-free and divergence-free state and exact agreement of
   the boundary values $\textbf{B}$ with observations $\textbf{B}_{\textrm{obs}}$ in regions where the elements of $\textbf{W}$ are
   greater than zero. For inconsistent boundary data the force-free and solenoidal conditions can
  still be fulfilled, but the surface term $\mathcal{L}_{\textrm{photo}}$ will remain finite. This results in some deviation of the
  bottom boundary data from the observations, especially in regions where $w_{\textrm{radial}}$ and $w_{\textrm{trans}}$ are
  small. The parameter $\nu$ is tuned so that these deviations do not exceed the local estimated measurement error.

   \end{itemize}
\section{Results}
In this work, we apply our extrapolation scheme to Milne-Eddington inverted vector magnetograph
data from the Synoptic Optical Long-term Investigations of the Sun survey (SOLIS). As a first
step for our work we remove non-magnetic forces from the observed surface magnetic field using
our spherical preprocessing procedure. The code takes $\textbf{B}_{\textrm{obs}}$ as improved boundary
condition.

SOLIS/VSM provides full-disk vector-magnetograms, but for some individual pixels the inversion
from line profiles to field values may not have been successful inverted and field data there
will be missing for these pixels (see Figure~\ref{fig1}). The different errors for the radial and
transverse components of $\textbf{B}_{\textrm{obs}}$ are taken account by different values for $w_{\textrm{radial}}$
and $w_{\textrm{trans}}$. In this work we used $w_{\textrm{radial}}=20w_{\textrm{trans}}$ for the surface preprocessed fields as
the radial component of $\textbf{B}_{\textrm{obs}}$ is measured with higher accuracy.

\begin{figure}
\centering
\includegraphics[viewport=1 1 595 525,clip,height=7.5cm,width=7.5cm]{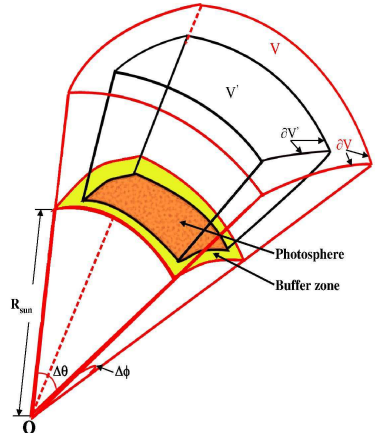}
\caption{Wedge-shaped computational box of volume $V$ with the inner physical
domain $V'$ and a buffer zone. O is the center of the Sun.}
\label{fig1b}
\end{figure}

We compute the 3D magnetic field above the observed surface region inside wedge-shaped computational
box of volume $V$, which includes an inner physical domain $V'$ and a buffer zone(the region outside
the physical domain), as shown in Figure~\ref{fig1b}. The physical domain $V'$ is a wedge-shaped volume,
with two latitudinal boundaries at $\theta_{\textrm{min}}=-26^{\circ}$ and $\theta _{\textrm{max}}=16^{\circ}$,
two longitudinal boundaries at $\phi_{\textrm{min}}=129^{\circ}$ and $\phi _{\textrm{max}}=226^{\circ}$, and two
radial boundaries at the photosphere ($r=1R_{\odot}$) and $r=1.75R_{\odot}$. Note that the longitude $\phi$ is
measured from the center meridian of the back side of the disk.
\begin{figure}
   \centering
   \mbox{
   \subfloat[28 March 2008 15:45UT]{\includegraphics[bb=30 30 560 560,clip,height=3.5cm,width=4.0cm]{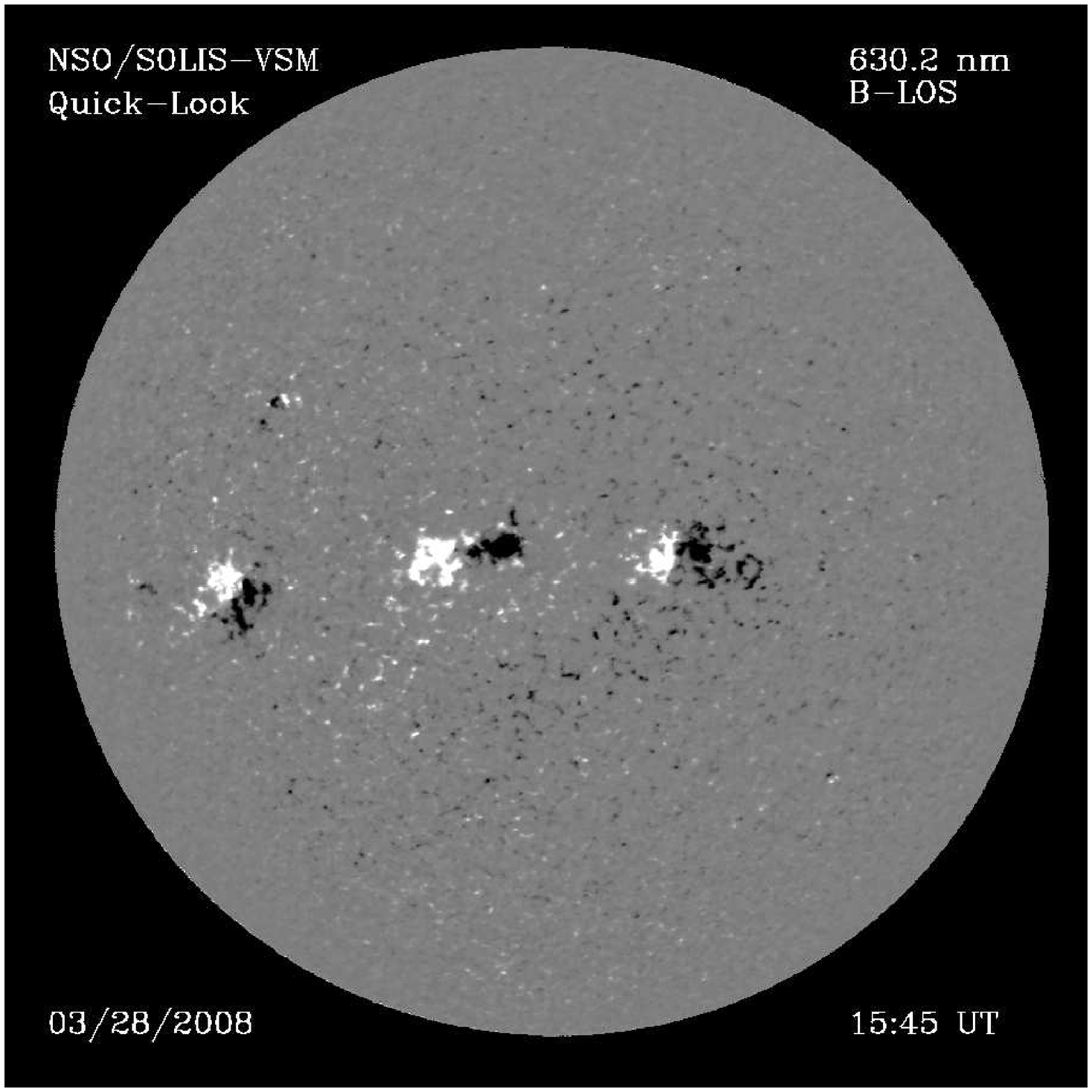}}\quad
   \subfloat[29 March 2008 15:48UT]{\includegraphics[bb=30 30 560 560,clip,height=3.5cm,width=4.0cm]{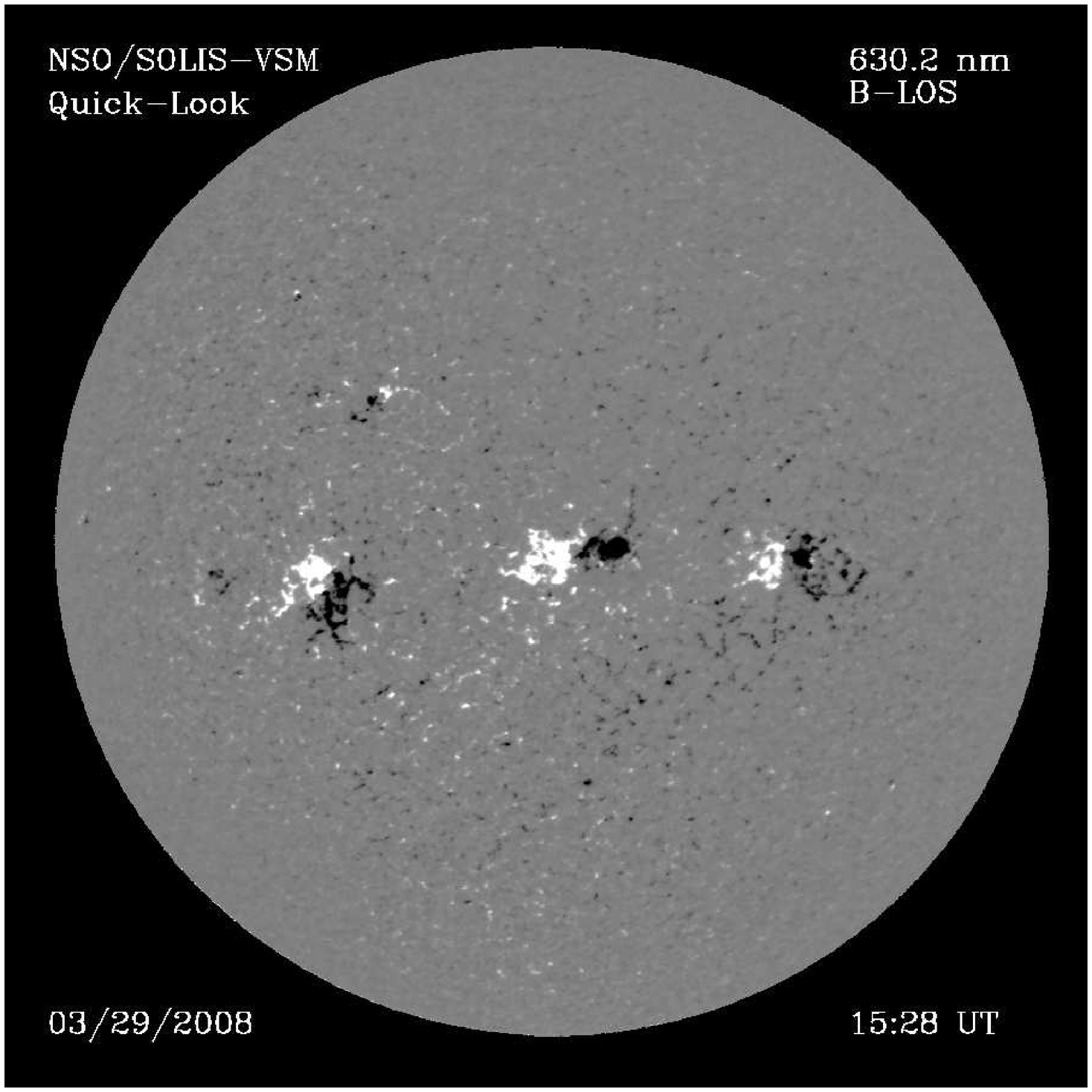}}\quad
   \subfloat[30 March 2008 15:47UT]{\includegraphics[bb=30 30 560 560,clip,height=3.5cm,width=4.0cm]{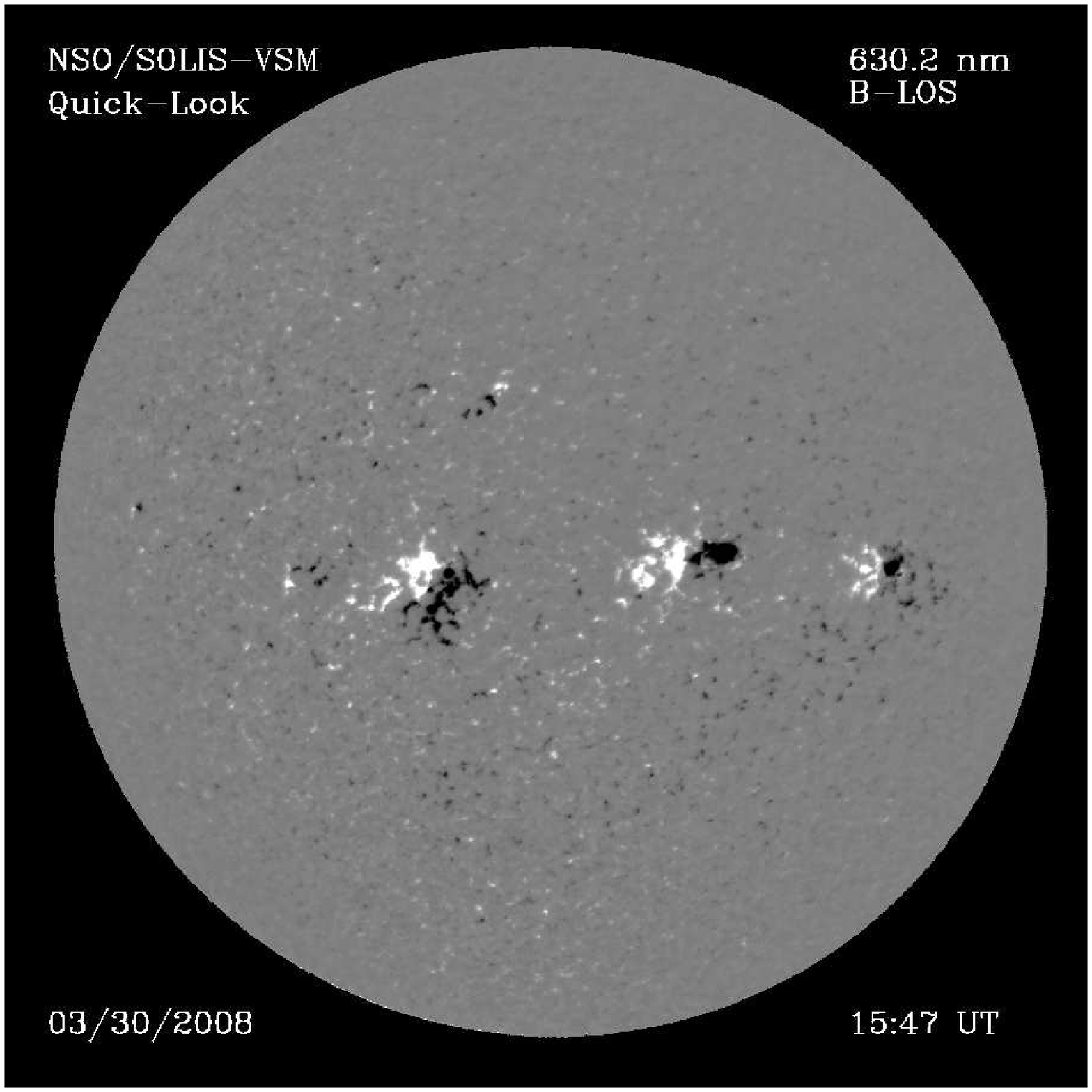}}}
   \mbox{
  \subfloat[28 March 2008 15:45UT]{\includegraphics[bb=175 100 387 298,clip,height=3.5cm,width=4.0cm]{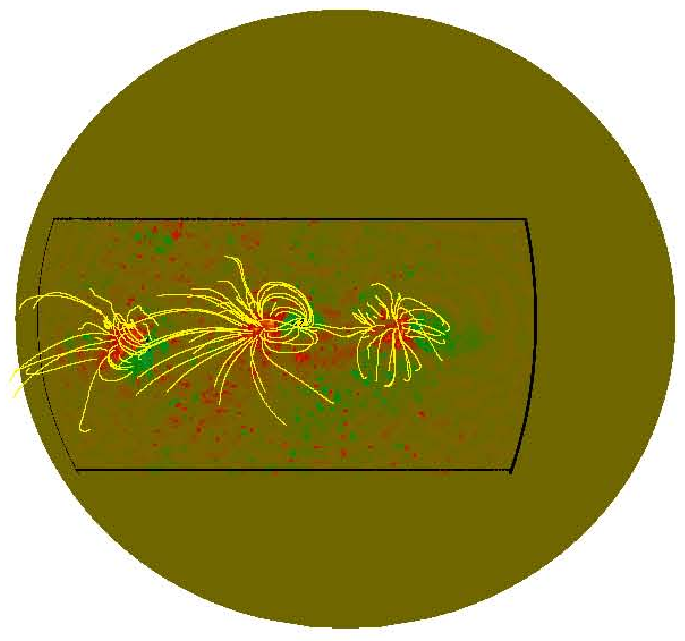}}\quad
  \subfloat[29 March 2008 15:48UT]{\includegraphics[bb=175 100 387 298,clip,height=3.5cm,width=4.0cm]{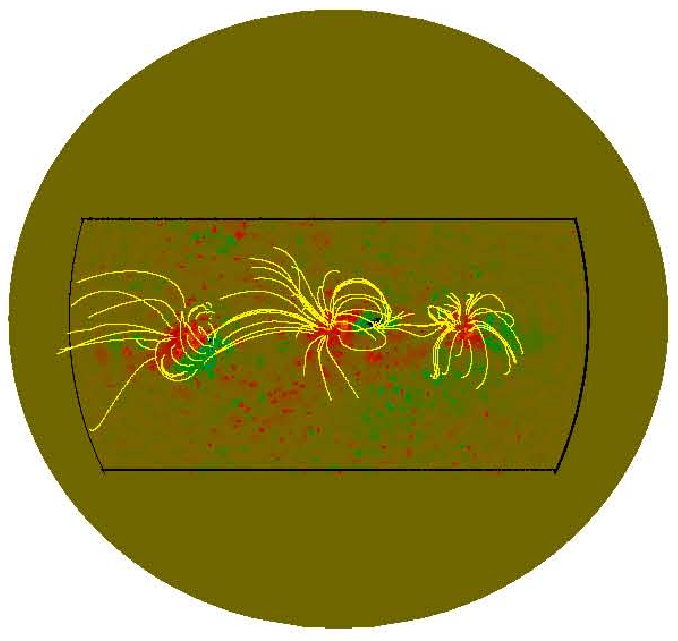}}\quad
  \subfloat[30 March 2008 15:47UT]{\includegraphics[bb=175 100 387 298,clip,height=3.5cm,width=4.0cm]{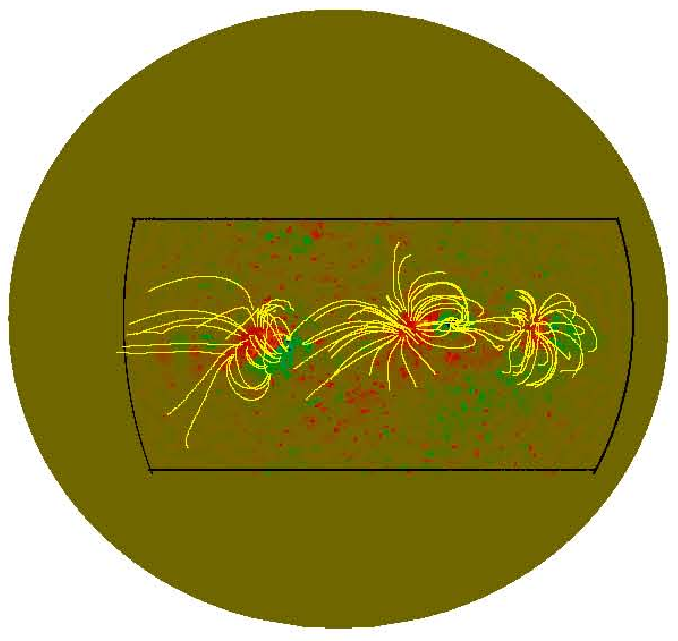}}}
  \mbox{
    \subfloat[28 March 2008 16:00UT]{\includegraphics[bb=33 50 550 530,clip,height=3.5cm,width=4.0cm]{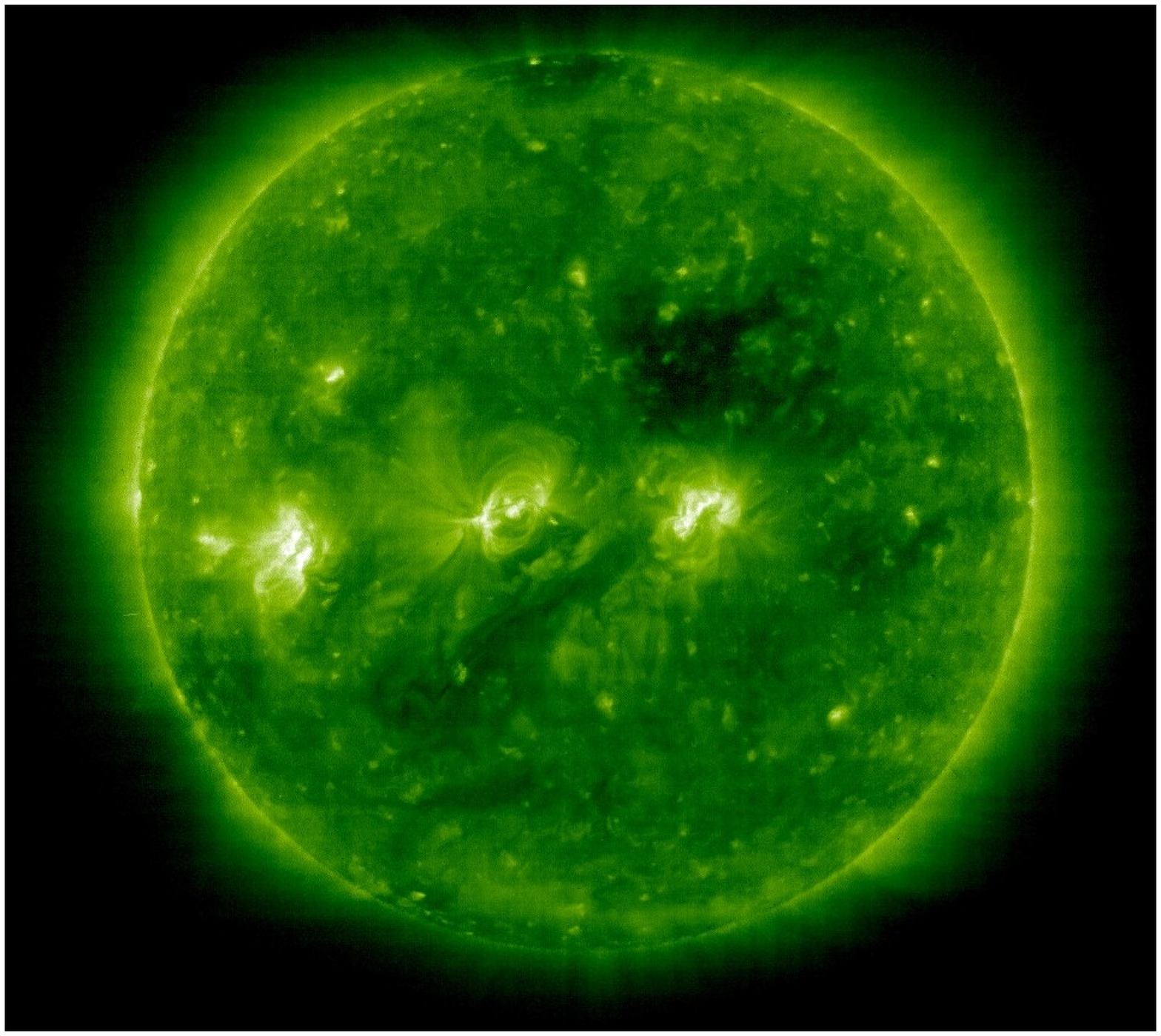}}\quad
   \subfloat[29 March 2008 15:48UT]{\includegraphics[bb=33 50 550 530,clip,height=3.5cm,width=4.0cm]{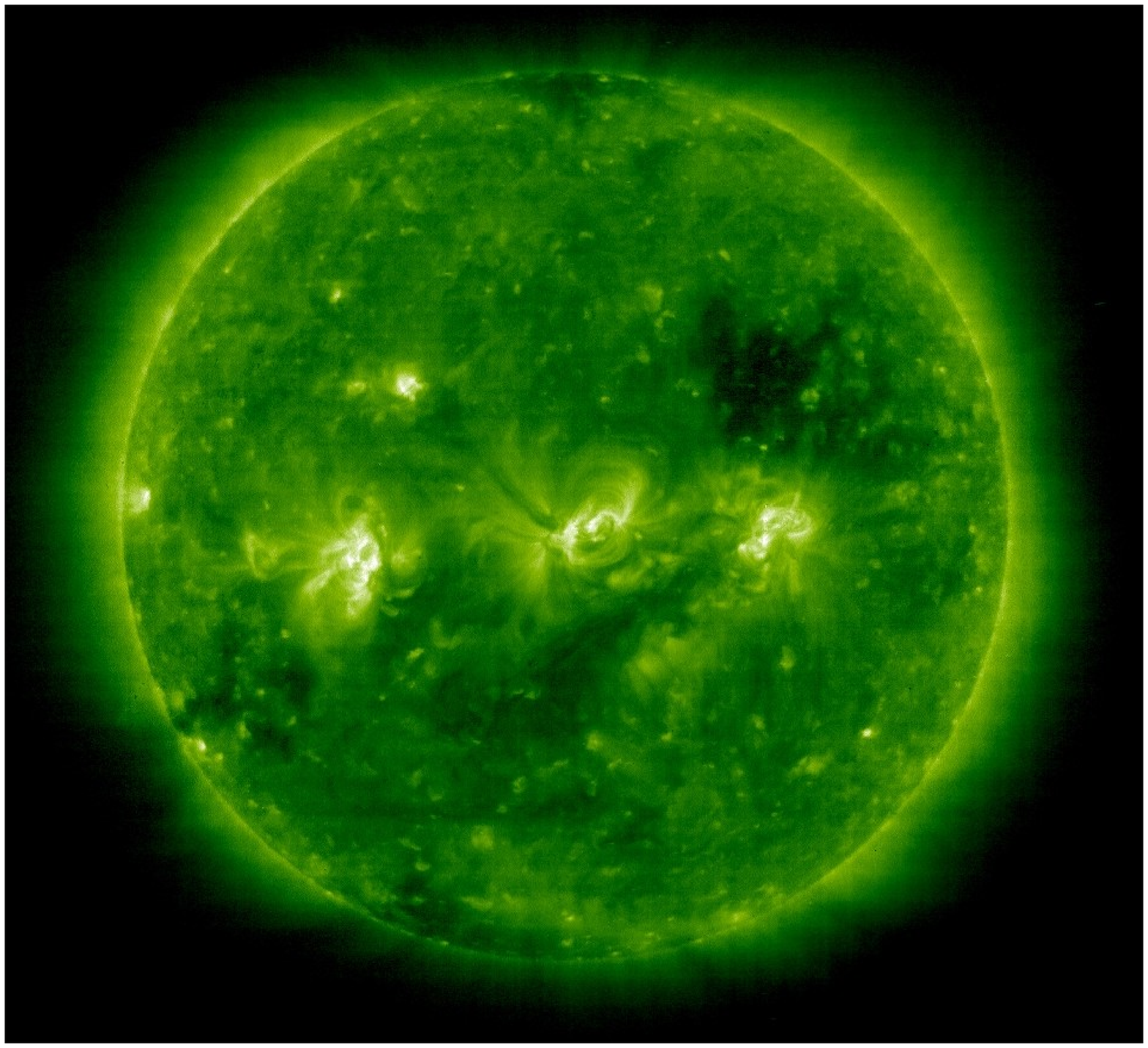}}\quad
   \subfloat[30 March 2008 15:48UT]{\includegraphics[bb=33 50 550 530,clip,height=3.5cm,width=4.0cm]{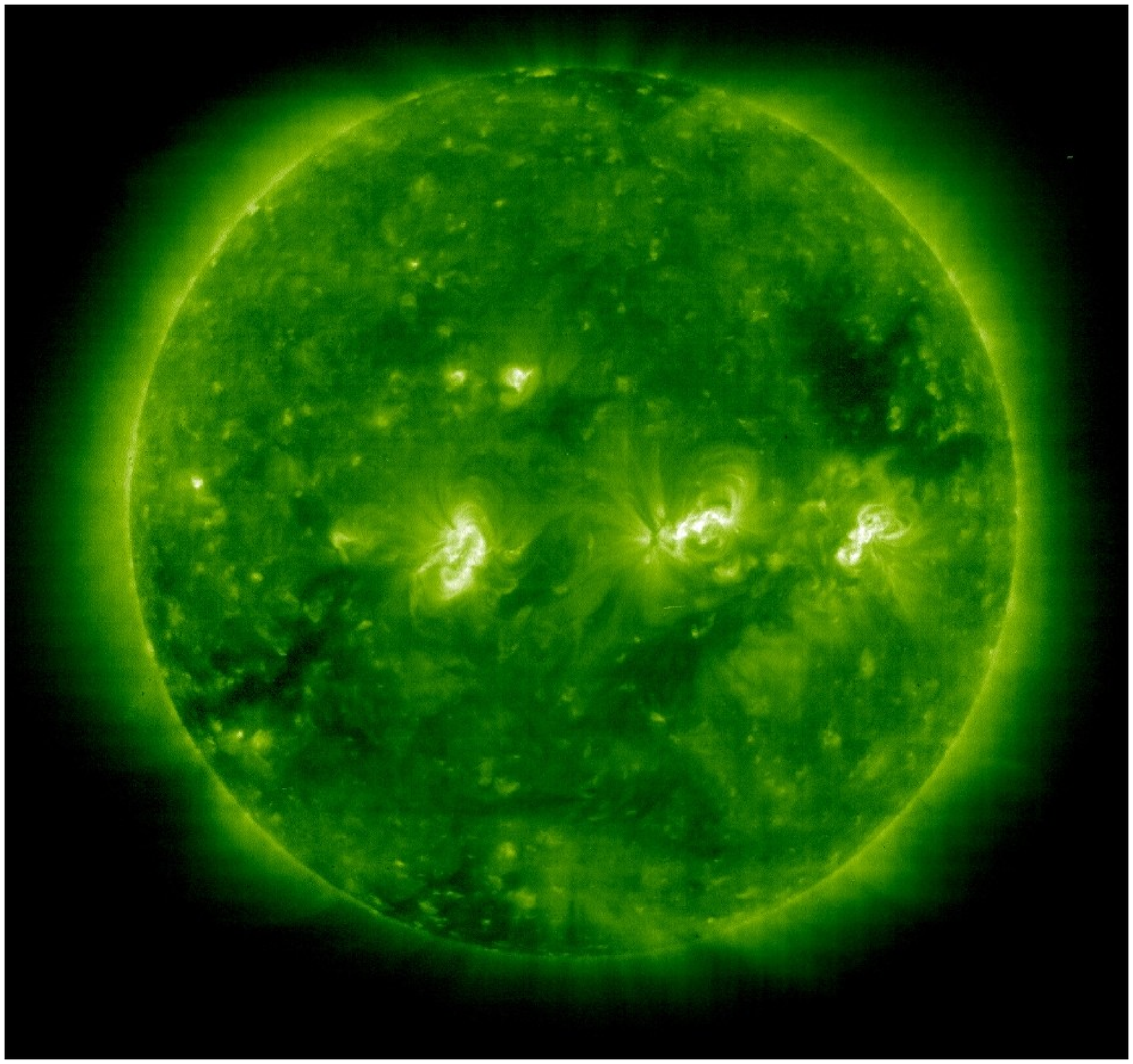}}}

   \caption{Top row: SOLIS/VSM magnetograms of respective dates.
   Middle row: Magnetic field lines reconstructed from magnetograms on the
   top panel.
   Bottom row: EIT image of the Sun at 195{\AA} on indicated dates.
   }
\label{fig2}
\end{figure}
\begin{figure}[htp!]
\begin{center}
  \subfloat[28 March 2008 15:45UT]{ \includegraphics[bb=172 147 345 235,clip,height=5.0cm,width=9.50cm]{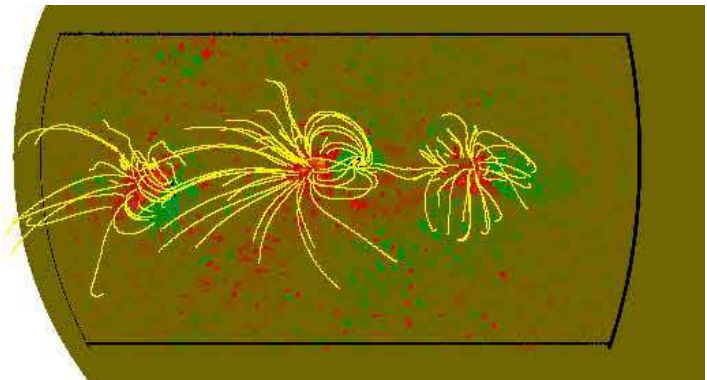}}\\
  \subfloat[29 March 2008 15:48UT]{ \includegraphics[bb=190 147 365 235,clip,height=5.0cm,width=9.50cm]{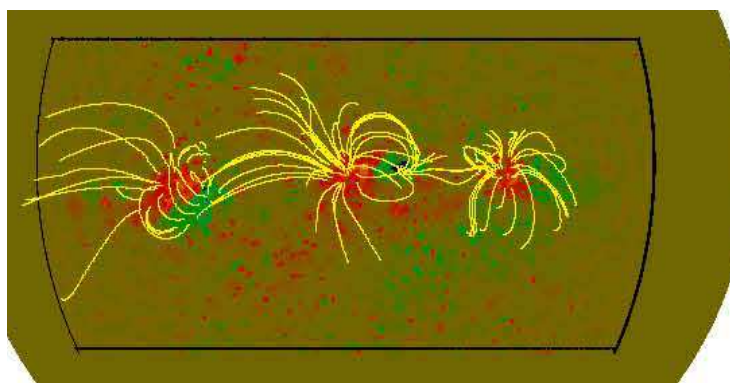}}\\
  \subfloat[30 March 2008 15:47UT]{ \includegraphics[bb=204 147 378 235,clip,height=5.0cm,width=9.50cm]{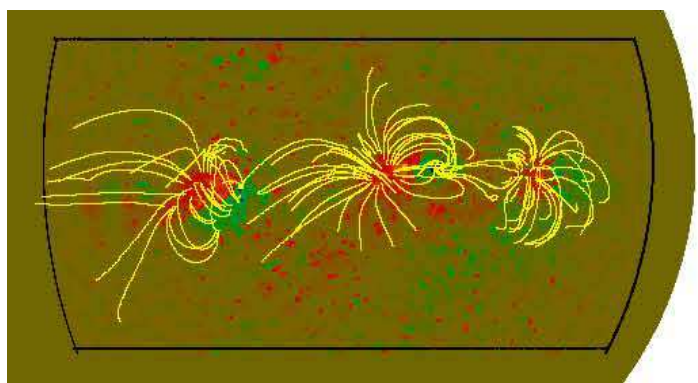}}
  \end{center}
     \caption{Some magnetic field lines plots reconstructed from SOLIS magnetograms using nonlinear force-free modelling.
The color coding shows $B_{r}$ on the photosphere.}
\label{fig3}
\end{figure}
We define $V'$ to be the inner region of $V$ (including the photospheric
boundary) with $\omega_{\textrm{f}}=\omega_{\textrm{d}}= 1$ everywhere including its six inner boundaries
$\delta V'$. We use a position-dependent weighting function to introduce
a buffer boundary of $nd = 10$ grid points towards the side and top boundaries
of the computational box, $V$. The weighting functions, $\omega_{\textrm{f}}$ and $\omega_{\textrm{d}}$ are chosen to
be unity within the inner physical domain $V'$ and decline to 0 with a cosine
profile in the buffer boundary region \cite{Wiegelmann04,Tilaye:2009}. The
framed region in Figure \ref{fig1} corresponds to the lower boundary of the
physical domain $V'$ with a resolution of $114\times 251$ pixels in the photosphere.

The middle panel of Figure~\ref{fig2} shows magnetic field line plots for three consecutive
dates of observation. The top and bottom panels of Figure~\ref{fig2} show the position of the three
active regions on the solar disk both for SOLIS full-disk magnetogram
\footnote{http://solis.nso.edu/solis\_data.html} and SOHO/EIT
\footnote{http://sohowww.nascom.nasa.gov/data/archive} image of the Sun
observed at 195{\AA} on the indicated dates and times. Figure~\ref{fig3}
shows some selected magnetic field lines from reconstruction from the SOLIS
magnetograms, zoomed in from the middle panels of Figure~\ref{fig2}. In each column
of Figure~\ref{fig3} the field lines are plotted from the same foot points to compare
the change in topology of the magnetic field over the period of the three days of observation.
In order to compare the fields at the three consecutive days quantitatively, we
computed the vector correlations between the three field configurations. The vector
correlation ($C_\mathrm{\textrm{vec}}$) \cite{Schrijver06} metric generalizes the standard correlation coefficient
for scalar functions and is given by
\begin{equation}
C_\mathrm{ \textrm{vec}}=  \frac{\sum_i \textbf{v}_{i} \cdot \textbf{u}_{i}}{\sqrt{\sum_i |\textbf{v}_{i}|^2} \sqrt{\sum_i
|\textbf{u}_{i}|^2} }\label{nine}
\end{equation}
where $\textbf{v}_{i}$ and $\textbf{u}_{i}$ are 3D vectors at grid point $i$. If the vector
fields are identical, then $C_{\textrm{vec}}=1$; if $\textbf{v}_{i}\perp \textbf{u}_{i}$ , then $C_{\textrm{vec}}=0$. The correlations
($C_\mathrm{\textrm{vec}}$) of the 3D magnetic field vectors of 28 and 30 March with
respect to the field on 29 March are $0.96$ and $0.93$ respectively. From these
values we can see that there has been no major change in the magnetic field
configuration during this period.
\begin{figure}[htp!]
\begin{center}
  \subfloat[28 March 2008 15:45UT]{ \includegraphics[bb=90 30 465 285,clip,height=6.50cm,width=10.50cm]{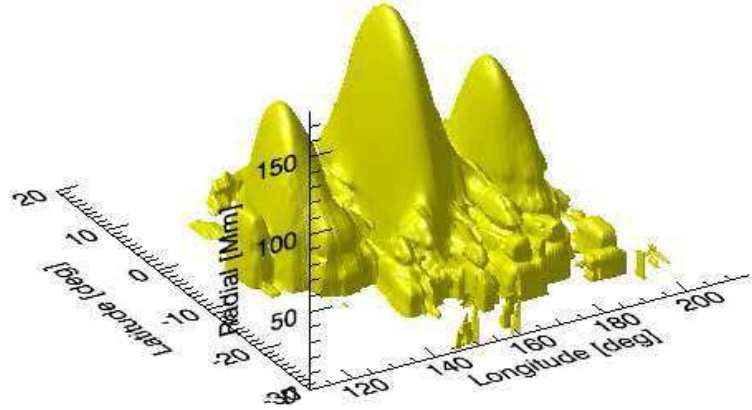}}\\
  \subfloat[29 March 2008 15:48UT]{ \includegraphics[bb=90 30 465 285,clip,height=6.50cm,width=10.50cm]{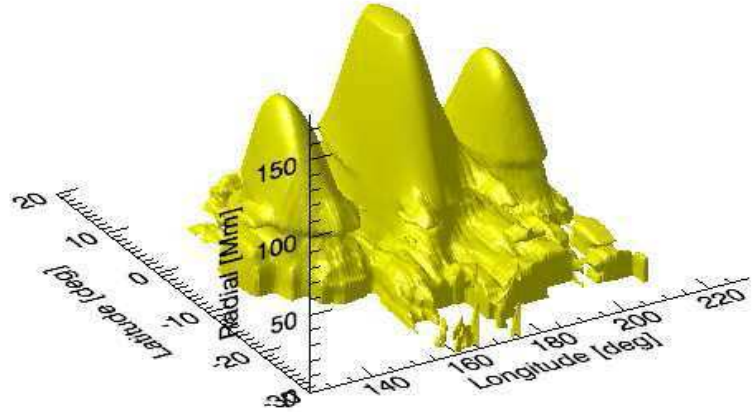}}\\
  \subfloat[30 March 2008 15:47UT]{ \includegraphics[bb=90 30 465 285,clip,height=6.50cm,width=10.50cm]{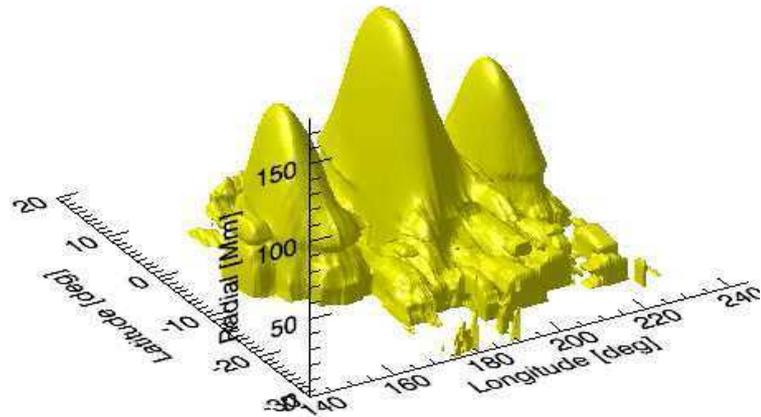}}
\end{center}
     \caption{Iso-surfaces (ISs) of the absolute NLFF magnetic energy density ($7.5\times10^{16}$\textrm{erg}) for the three consecutive dates
     computed within the entire computational domain.}
\label{fig4}
\end{figure}

We also compute the values of the free magnetic energy estimated from the excess
energy of the extrapolated field beyond the potential field satisfying the same
$\textbf{B}_{\textrm{obs}}\cdot\hat{r}$ boundary condition. Similar estimates have
been made by \inlinecite{Regnier} and \inlinecite{Thalmann} for single active regions observed at
other times. From the corresponding potential and force-free magnetic field, $\textbf{B}_{\textrm{pot}}$
and $\textbf{B}$, respectively, we can estimate an upper limit to the free magnetic energy associated with coronal
currents
\begin{equation}
E_\mathrm{free}=E_\mathrm{\textrm{nlff}}-E_\mathrm{\textrm{pot}}=
\frac{1}{8\pi}\int_{V'}\Big(B_{\textrm{nlff}}^{2}-B_{\textrm{pot}}^{2}\Big)r^{2}\textrm{sin}\theta dr d\theta d\phi. \label{ten}
\end{equation}

\begin{center}
\begin{table}
\begin{tabular}{cccc}
 \hline
 Date & $E_{\textrm{nlff}}(10^{32}\textrm{erg})$&$E_\mathrm{\textrm{pot}}(10^{32}\textrm{erg})$& $E_\mathrm{free}(10^{32}\textrm{erg})$\\
\hline
28 March 2008 &$57.34$&$53.89$&$3.45$\\
29 March 2008 &$57.48$&$54.07$&$3.41$\\
30 March 2008 &$57.37$&$53.93$&$3.44$\\
\hline
\end{tabular}
\caption{The magnetic energy associated with extrapolated NLFF field configurations for the three particular dates.}
\label{table1}
\end{table}
\end{center}
The computed energy values are listed in Table~\ref{table1}. The free energy on all three
days is about $3.5\times10^{32}\textrm{erg}$. The magnetic energy associated with the
potential field configuration is about $54\times10^{32}\textrm{erg}$. Hence $E_{\textrm{nlff}}$ exceeds $E_{\textrm{pot}}$ by only 6$\%$.
Figure~\ref{fig4} shows Iso-surface plots of magnetic energy density in the volume above the active regions.
There are strong energy concentrations above each active region. There were no major changes in the magnetic
energy density over the observation period and there was no major eruptive phenomenon during those three days
in the region observed.

In our previous work\cite{Tilaye:2010}, we have studied the connectivity between two neighbouring active
regions. In this work with an even larger field of view, the three ARs share a decent amount of magnetic flux compared to their
internal flux from one polarity to the other (see Figure~\ref{fig3}). In terms of the electric current they are
much more isolated. In order to quantify these connectivities, we have calculated the magnetic flux and the electric
currents shared between active regions. For the magnetic flux, \textit{e.g.}, we use
\begin{equation}
\Phi_{\alpha\beta}=\sum_{i}|\textbf{B}_{i}\cdot\hat{r}|R^{2}_{\odot}\textrm{sin}(\theta_{i})\Delta\theta_{i}\Delta\phi_{i}\label{ten}
\end{equation}
where the summation is over all pixels of $\mbox{AR}_{\alpha}$ from which the field line
ends in $\mbox{AR}_{\beta}$ or $i\in\mbox{AR}_{\alpha}\|\,\mbox{conjugate footpoint}(i)\in\mbox{AR}_{\beta}$. The indices $\alpha$
and $\beta$ denote the active regions and the index number $1$ corresponds to AR $10989$, $2$ to AR $10988$ and $3$ to
AR $10987$ of Figure~\ref{fig1}. For the electric current we replace the magnetic field, $\textbf{B}_{i}\cdot\hat{r}$, by the
vertical current density $\textbf{J}_{i}\cdot\hat{r}$ in Equation (\ref{ten}). Whenever the end point
of a field line falls outside (blue rectangles in Figure~\ref{fig1}) of the three ARs, we categorize it as ending elsewhere. Both
Table~\ref{table2} and~\ref{table3} show the percentage of the total magnetic flux and
electric current shared between the three ARs. So, for example first column of Table~\ref{table2} shows
that $56.37\%$ of positive polarity of $\mbox{AR}_{1}$ is connected to negative polarity of $\mbox{AR}_{1}$; line 2 shows
that $13.66\%$ of positive/negative polarity of $\mbox{AR}_{1}$ is connected to positive/negative polarity of $\mbox{AR}_{2}$, and
line 3 shows that there are no field lines $(0\%)$ connecting positive/negative polarity of $\mbox{AR}_{1}$ with positive/negative
polarity of $\mbox{AR}_{3}$. The same technique applies for Table~\ref{table3} too.
The three active regions are magnetically connected but much less by electric currents.
\begin{table}
\begin{tabular}{rcclrclrcc}
\cline{2-10}
&& $28th$& &&$29th$&&&$30th$& \\
\cline{2-10}
$\Phi_{\alpha\beta}$ &$\alpha=1$ & $2$ & $3$&$\alpha=1$ & $2$ & $3$&$\alpha=1$ & $2$ & $3$\\
\cline{1-10}
\multicolumn{1}{r}{$\beta=1$}& $56.37$&$5.59$&$0.00$&   $56.50$&$5.48$&$0.00$&   $56.50$&$5.48$&$0.00$\\
\multicolumn{1}{r}{$2$}& $13.66$&$81.12$&$1.43$&  $13.66$&$81.22$&$1.43$&    $13.66$&$81.22$&$2.22$\\
\multicolumn{1}{r}{$3$}& $0.00$&$0.48$&$71.47$&    $0.00$&$0.48$&$71.80$&    $0.00$&$0.48$&$71.80$\\
\multicolumn{1}{r}{Elsewhere} & $29.97$&$12.82$&$27.10$&   $29.84$&$12.82$&$26.77$&    $29.84$&$12.82$&$25.98$\\
\cline{1-10}
\end{tabular}
\caption{The percentage of the total magnetic flux shared between the three ARs. $\Phi_{11}$, $\Phi_{22}$ and $\Phi_{33}$
denote magnetic flux of AR 10989(left), AR 10988(middle) and AR 10987(right) of Figure~\ref{fig1}, respectively. }
\label{table2}
\end{table}
\begin{table}
\begin{tabular}{rcclrclrcc}
\cline{2-10}
&& $28th$& &&$29th$&&&$30th$& \\
\cline{2-10}
$I_{\alpha\beta}$ &$\alpha=1$ & $2$ & $3$&$\alpha=1$ & $2$ & $3$&$\alpha=1$ & $2$ & $3$\\
\cline{1-10}
\multicolumn{1}{r}{$\beta=1$}& $82.47$&$0.19$&$0.00$&    $86.36$&$0.19$&$0.00$&     $94.16$&$0.19$&$0.00$\\
\multicolumn{1}{r}{$2$}& $0.65$&$85.25$&$1.42$&    $0.65$&$85.25$&$1.42$&    $0.65$&$85.25$&$3.55$\\
\multicolumn{1}{r}{$3$}& $0.00$&$0.38$&$82.27$&     $0.00$&$0.38$&$82.27$&      $0.00$&$0.38$&$82.27$\\
\multicolumn{1}{r}{Elsewhere} & $16.88$&$14.18$&$16.31$&    $12.99$&$14.18$&$16.31$&      $5.19$&$14.18$&$14.18$\\
\cline{1-10}
\end{tabular}
\caption{The percentage of the total electric current shared between the three ARs. $I_{11}$, $I_{22}$, and $I_{33}$
denote electric current of AR 10989(left), AR 10988(middle) and AR 10987(right) of Figure~\ref{fig1}, respectively. }
\label{table3}
\end{table}
\section{Conclusions}
\label{sect:disc}
We have investigated the coronal magnetic field associated with three ARs 10987,
10987, 10989, on  28, 29 and 30 March 2008 by analysing SOLIS/VSM data. We have
used an optimization method for the reconstruction of nonlinear force-free
coronal magnetic fields in spherical geometry by restricting the code to limited
parts of the Sun \cite{Wiegelmann07,Tilaye:2009,Tilaye:2010}. The code was modified
so that it allows us to deal with lacking data and regions with poor signal-to-noise
ratio in a systematic manner\cite{Wiegelmann10,Tilaye:2010}.

We have studied the time evolution of magnetic field over the period of three
days and found no major changes in topologies as there was no major eruption
event. The magnetic energies calculated in the large wedge-shaped computational
box above the three active regions were not far apart in value. This is the first study
which contains three well separated ARs in our model. This was made possible by the use
of spherical coordinates and it allows us to analyse linkage between the ARs. The active
regions share a decent amount of magnetic flux compared to their internal flux from one
polarity to the other. In terms of the electric current they are much more isolated.

\section*{Acknowledgements} SOLIS/VSM vector magnetograms are produced cooperatively by NSF/NSO and NASA/LWS.
The National Solar Observatory (NSO) is operated by the Association of Universities for Research in Astronomy,
Inc., under cooperative agreement with the National Science Foundation. Tilaye Tadesse Asfaw acknowledges a fellowship of
the International Max-Planck Research School at the Max-Planck Institute for Solar System Research and the work of T.
Wiegelmann was supported by DLR-grant $50$ OC $0501$.

\bibliographystyle{spr-mp-sola}
\bibliography{paper_2010_bibtex}

\end{article}
\end{document}